# Regional modeling of surface and sub-surface dynamics in the Bay of Bengal using Modular Ocean Model

Siddhesh Tirodkar[a], Mousumi Sarkar[a], Rajesh Chauhan[a], Manasa R. Behera[a,b], Sridhar Balasubramanian[a,c]

[a]IDP in Climate Studies, IIT Bombay, Powai, Mumbai 400076
[b]Department of Civil Engineering, IIT Bombay, Powai, Mumbai 400076
[c]Department of Mechanical Engineering, IIT Bombay, Powai, Mumbai 400076

**Abstract:**

Regional dynamics of Bay of Bengal is studied using Open Boundary Condition (OBC) in Modular Ocean Model (MOM) to understand the effect of primary and secondary mesoscale features on various bulk ocean products and turbulent fluxes. A horizontal resolution of $0.1°$ is adopted to resolve a range of mesoscale features. In order to parametrize the vertical mixing, K-Profile Parametrization (KPP) scheme is implemented.

The results show successful implementation of OBC in a regional domain with exchange of mass and energy and conservation of mass at the boundaries. The Sea Surface Temperature (SST) and Sea Surface Salinity (SSS) are validated with SODA reanalysis, which are found to be in good agreement. The Mixed Layer Depth (MLD) pattern is very well represented, albeit the magnitude is slightly under-predicted in comparison with SODA reanalysis.

Additionally, analysis of flow energetics reveal regional differences in turbulent kinetic energy (K), production flux (P), buoyancy flux (B), and dissipation ($\varepsilon$). Our results clearly reveal the presence of inverse energy-cascade in the southern Bay of Bengal, wherein energy flows back into mean flow structures from the turbulent eddies. The re-energization of mean flow structures is likely to allow the large-scale circulation to persist for longer periods, thereby modifying the local dynamics. Further analysis shows that the B term is the major source of turbulence production in the north and central Bay of Bengal regions. These results convey the various mechanisms by which energy is produced and transferred in different regions of Bay of Bengal.

# 1  INTRODUCTION

The surface dynamics, vertical structure and sub-surface processes in the ocean play a crucial role in modulating weather and climate. The air-sea fluxes define the response of the ocean processes to sea surface temperature (SST), sea surface salinity (SSS), and the extent of vertical mixing or mixed layer depth (MLD). The Bay of Bengal (BoB) experiences high input of fresh water flux and seasonal wind reversal. Dominant south-westerly wind during summer monsoon (June–September) and north-easterly wind during winter monsoon (November–February) show presence of turbulent fluxes reversal with upper ocean currents (Potemra et al., 1991; Tsai et al., 1992; Perigaud and Delecluse, 1992; McCreary et al., 1993; Masumoto and Meyers,1998). The Bay of Bengal Monsoon Experiment (BOBMEX) ($17^O$ 30' N, $89^O$ 00' E) (Bhat et al., 2001; Bhat, 2003) data analysis reveals the significant contribution of horizontal heat advection over vertical heat advection to MLD due to presence of strong barrier layer. Seasonal wind patterns and associated current reversal with ocean properties are analysed with model based studies by Schott and McCreary (2001); Shankar et al. (2002); Thompson et al. (2006). Chatterjee et al. (2017) studied the effect of Equatorial Kelvin waves (EKW) on circulations in BoB and observed that the Andaman and Nicobar Islands not only influence the circulation within Andaman Sea, but also significantly alter the circulation in interior bay and along the east coast of India. The study performed by Suneet et al. (2019) found that air-sea fluxes play a dominant role in the seasonal evolution of SST in the north Indian Ocean and the contribution of horizontal advection, vertical entrainment and diffusion processes is small. Francis et al. (2020) analysed structure and variability of undercurrents in East India Coastal Current (EICC) and the mechanisms of their formation. It is suggested that the undercurrents, observed below the EICC, were a part of distinct anticyclonic eddies. Interaction of westward propagation of such eddies with EICC weakened the strong surface flow.

The large variability of basin features incorporate many primary and secondary mesoscale and sub-mesoscale processes in the Bay of Bengal basin. Most of the fine scale process studies are observational in nature (Prasanna Kumar et al., 2004; Nuncio and Kumar, 2012; Cui et al., 2016; Cheng et al., 2018; Roman-Stork et al., 2019; Gulakaram et al., 2018, 2020) and have limitation in terms of spatial data. Few modeling studies with Regional Ocean Modeling System (ROMS) and Australian Community Ocean Model (ACOM) have been carried out by Kurien et al. (2010); Francis et al. (2013); Mukherjee et al. (2018); Effy et al. (2020); Francis et al. (2020), which suggest that the effect of fine scales do not impact the ocean properties - a result that is contradictory to physics based observation and theory. A basin-wide modeling approach is generally preferred, as opposed to global modeling, for studying the effect of fine scale processes on ocean dynamics Mukherjee et al. (2018). Available research studies have concentrated on the dynamics and vertical structure of ocean by resolving the mesoscale processes using sponge boundary conditions (Kurian and Vinayachandran, 2007; Chatterjee et al., 2013; Behara and Vinayachandran, 2016). The study by Srivastava et al. (2018) has used open boundary condition with prescription of data at boundary, but using MITgcm.

Open boundary conditions are the preferred method for performing regional domain study at fine resolution. Marsaleix et al. (2006) reviews the usual OBCs for coastal ocean models and proposes a complete set of open boundaries based on stability criteria, mass and energy conservation arguments, and the ability to enforce external information. A universal OBC is not possible and its implementation is domain dependent. Moreover, OBC has been used for ocean modeling of the Atlantic basin, but, there exists no research work with open boundary condition in Modular Ocean Model (MOM) for the Indian Ocean basin. Therefore, our study is first of its kind, where the functionality of open boundary condition is tested for modeling Bay of Bengal dynamics using MOM5.

The present study focuses on seasonal variability in BoB, by looking at the surface parameters, vertical structure, and energetics analysis. Section 2 includes details of the model set-up, data set used and experiments performed. Section 3 focuses on seasonal behaviour observed in Bay of Bengal such as the surface currents, radiation heat budget, mixed layer depth, model temperature and salinity comparison with reanalysis data. The seasonal variability of turbulent kinetic energy (K), production flux (P), buoyancy flux (B), viscous dissipation ($\epsilon$) and advection of K is discussed in section 4. The conclusions are presented in section 5.

## 2 MODEL, DATA AND METHODOLOGY

The model selected to perform this study is Modular Ocean Model (MOM) version 5, designed by NOAA's Geophysical Fluid Dynamics Laboratory (GFDL). MOM is 3-dimensional hydro-static generalized level coordinate numerical model evolved from the base model, developed by Bryan and Cox (Bryan and Cox, 1967, 1972).

Fig. 1 shows the model domain in the Indian Ocean selected for this study. The latitudinal extent is 10° S to 25° N, which is selected to account for cross-equatorial dynamics. The longitudinal extent is 75° E to 100° E and covers only the Bay of Bengal region. For analysis we are only considering 3° N to 24° N in latitude and 78° E to 94° E in longitudes to avoid eastern island and equatorial cross-flows. An Arakawa B-grid scheme is used to compute velocities at the corner of the cell and tracer at the centre of the cell. Uniform horizontal grid resolution of 1/10$^{th}$ degree (0.1°, ~11 km) is maintained in zonal and meridional direction. The model has non-uniform vertical grid with 5 m uniform grid in upper 60 m depth near surface and then gradually increases till maximum depth of 5000 m. The bathymetry used in the study domain is the modified Indian Ocean ETOPO2V2 (Sindhu et al., 2007).

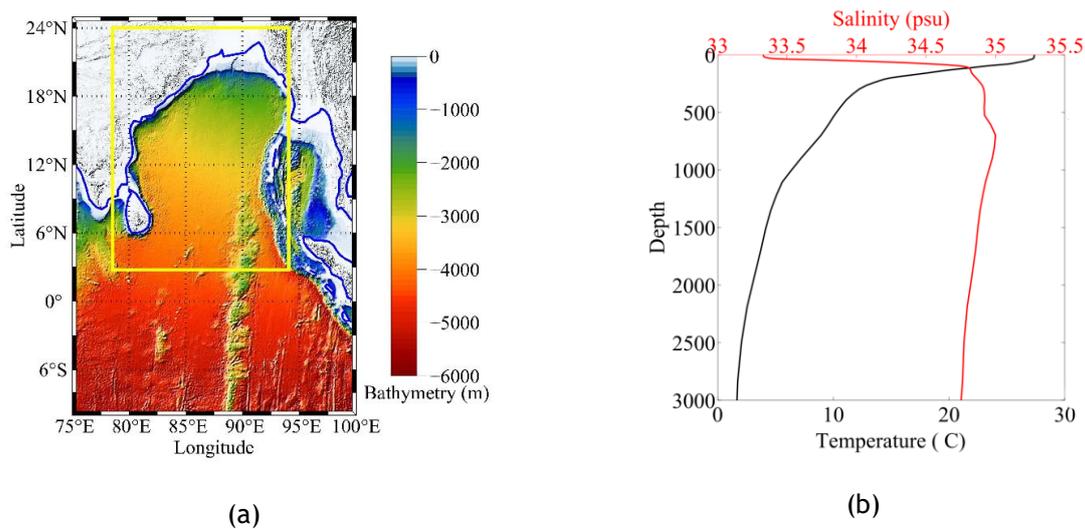

(a) (b)

Figure 1: (a)Study domain. Yellow rectangle shows the domain analysed (b)Initial temperature and salinity vertical profile up to 3000 m depth (temperature (black line) and salinity (red line) at the center of domain)

Since the study domain is part of BoB, it has few boundaries interacting with the open ocean. Different type of open boundary conditions available in MOM5 which are well described in Herzfeld et al. (2011). Literature confirms that selection of OBC is domain specific and in general a

"universal" artificial boundary condition is not available (Jensen, 1998). The domain extent is mainly in the north part of the Indian Ocean; hence, the domain is bounded by land from north and has three open boundaries: the entire south boundary, a part of west, and east boundary. These open boundaries exchange mass, momentum and heat with the outer ocean basin to maintain balance of ocean properties across the domain boundary. Our model runs show that for the BoB domain under consideration in our study, the *Orlanski* condition for tracers is the most appropriate out of the rest radiation BC. Hence, we proceeded with this boundary condition in all our simulation runs.

Fig. 1 shows the location of the domain analysed. All coastal boundaries in the domain represent land-ocean interaction. The north wall of the domain behaves as solid wall. It is usually assumed that the open boundaries should be largely driven by external incoming waves and thus a restoring term towards a forcing variable may be added to OBC (Marsaleix et al., 2006), thus the domain is prescribed with sea surface height data (TOPEX data; Fu et al.1994) at the boundaries to consider effect of incoming and outgoing waves. Temperature and salinity tracers are computed at the boundary with the implementation of "Orlanski" radiation condition as described in Orlanski(1976). The implementation of Orlanski radiation condition is as shown in Eq.1.

$$\eta(t+1, x_B) = \frac{(1-\mu)\eta(t-1,x_B) + 2\mu\eta(t,x_{B-1})}{1+\mu} \quad (1)$$

here, η is any tracer; μ is a dimensionless parameter which takes the following values:

$$\mu = \begin{cases} 1 \text{ if } \alpha \geq 1 \\ \alpha \text{ if } 0 < \alpha < 1 \\ 0 \text{ if } \alpha \leq 0 \end{cases}$$

where $\alpha = \frac{\eta(t-1,x_{B-1}) - \eta(t+1,x_{B-1})}{\eta(t+1,x_{B-1}) + \eta(t-1,x_{B-1}) - 2\eta(t,x_{B-2})}$. $c = \alpha \frac{\Delta x}{\Delta t}$ where, Δx is horizontal grid spacing and Δt is time step. The sub-script B, B-1, B-2 denote the grid points. Boundary points are marked with a capital B. The first point beyond the boundary outside the domain is B+1, the first internal point in the domain is B-1. The boundary and interior grid points, used in Arakawa B grid, are explained in Herzfeld et al. (2011); Griffies (2012).

Relaxation of ocean variable towards external data is implemented for surface elevation and tracer with following formulation in Eq.2.

$$T(t+1, x_B) = (T_0 - T(t, x_B))\frac{\Delta t}{\tau_f} \qquad (2)$$

where, T is tracer, relaxation time $\tau_f$ depends on the flow direction near the boundary. $\tau_f^{in}$ and $\tau_f^{out}$ can be specified in the name-list for each tracer and boundary separately. In our study tracers are prescribed at the boundaries with inflow relaxation time, $\tau_f^{in}$ of 1 hour (strong) and outflow relaxation time $\tau_f^{out}$ of 1 month to avoid shocks when changing from inflow to outflow situation.

The imperfect model parametrizations and inaccuracies in the heat and freshwater flux forcing data generally result into systematic model bias. Relaxing the data for temperature or salinity or both tracers is done, in order to prevent systematic drift in ocean variables (see for e.g.Cox and Bryan (1984); Shaji et al. (2003); Large and Yeager (2004); Thompson et al. (2006); Griffies et al. (2009); Francis et al. (2013); Courtois et al. (2017); Srivastava et al. (2018); Rahaman et al. (2019); Francis et al. (2020)). Hence, temperature and salinity are relaxed towards Simple Ocean Data Assimilation (SODA) pentad data with a timescale of 30 days. The relaxation of temperature and salinity is done with a flux correction method explained by Griffies (2003). The flux correction method considers both model and prescribing data for relaxation.

The model MOM5 employs depth-based z-coordinate, Boussinesq approximation. The model include provision of $f$-plane and β-plane approximation to consider effect of Coriolis force. Since the model domain is close to the equator, the β- plane approximation is used. For shortwave penetration into the upper ocean, the chlorophyll based scheme (Morel and Antoine, 1994) has been used. The model uses K-profile parameterization (KPP) to compute the vertical mixing. KPP with background vertical viscosity =$10^{-4}$ $m^2s^{-1}$; background diffusivity =$10^{-5}$ $m^2s^{-1}$; constant for pure convection = 1.8 (Large et al., 1994); critical Richardson number =0.25 (Kundu and Cohen,1990) is used. Horizontal mixing scheme used is Laplacian and Biharmonic friction scheme and Smagorinsky coefficient = 2.0 is considered (Chassignet and Garraffo, 2001).

The model is initialized with 3-dimensional profile of temperature and salinity. The model simulation starts from the state of rest with the annual averaged climatological temperature and salinity profile taken from World Ocean Atlas (WOA18) (Locarnini et al., 2018; Zweng et al., 2019).

Fig. 1(b) shows vertical profiles of temperature and salinity till 3000 m depth at the centre of the domain. These vertical profiles clearly show the existence of thermocline.

Table.1 Details of forced variable.

| Field Name | Data Source | Frequency | Reference |
|---|---|---|---|
| Initial Temperature (°C) | WOA18 | Annual | Locarnini et al. (2018) |
| Initial Salinity (psu) | WOA18 | Annual | Zweng et al. (2019) |
| Wind stress (N/m2) | SODA | Pentad | Carton et al. (2000) |
| Shortwave (W/m2) | WHOI | Daily | Yu et al. (2008) |
| Long Wave (W/m2) | WHOI | Daily | Yu et al. (2008) |
| Sensible Heat (W/m2) | WHOI | Daily | Yu et al. (2008) |
| Evaporation Rate (cm/yr) | WHOI | Daily | Yu et al. (2008) |
| Precipitation Rate (mm) | TRMM | Daily | https://trmm.gsfc.nasa.gov/ |
| Chlorophyll (mgm-3) | MODIS | Daily | Hu et al. (2012) |
| SSH (cm) | TOPEX | Weekly | Fu et al. (1994) |

Present study considers the effect of wind and solar forcing together, hence the parameters used in the study includes wind stress in zonal and meridional directions, shortwave radiation, long wave radiation, sensible heat, evaporation rate and precipitation rate. Table 1 gives more information about the forcing parameter used including field name, data source, frequency and reference. All the dataset are available at http://apdrc.soest.hawaii.edu. The data time-span is limited for duration from January 2003 to December 2012.

SODA reanalysis data (Carton et al., 2018), NIOA monthly Climatology (Chatterjee et al., 2012), WHOI data and Argo float data is used to validate model output. The bias between validation data and model is considered to analyse the accuracy of model simulation and ocean physics. Root-mean-square error (RMSE) (Chai and Draxler, 2014) is used as statistical measure to check the strength of the relationship between reanalysis and model output. The low value of RMSE indicates high accuracy of model results.

The forcing data is re-gridded to model grid with the help of model's in-built function. The model simulations are performed by converting the forcing data to climatology. The spin-up period of model simulation is considered when the total kinetic energy reaches an equilibrium state. In this study, the spin-up time is found to be ~ 5 years. A bulk run is performed for a period of 10 years including spin-up using the computed climatology data (for data from 2003-2012). Following this, controlled simulated runs are performed with instantaneous data for 5 years (for data from 2003-2007).

## 3   SEASONAL VARIABILITY OF MODEL OUTPUTS

Analysis is done for different seasons which are defined as DJF (Dec-Jan-Feb), MAM (Mar-Apr-May), JJAS (Jun-Jul-Aug-Sep), and ON (Oct-Nov) based on the predominant wind directions. The response in ocean currents due to the variations in seasonal wind reversal at (for) selected model parametrizations is analysed. All the results are averaged over 5 years (of instantaneous data simulation) unless it is explicitly mentioned. In this section, we discuss results validation of 0.1° horizontal resolution with reanalysis dataset.

### 3.1   Variability of currents

The open boundary condition mainly focuses on the inflow and outflow of tracers and wave or current propagation in and out of the domain. The prescription of TOPEX sea surface height data at the boundary ensures inflow of barotropic, baroclinic and other waves generated near the open boundaries and in IO basin, outside the study domain. The designed parametrization ensures conservation of mass inside the domain. It is also noticed that problem of emptying basin occurs if the sea surface height is not prescribed at the open boundary.

Inflow and outflow of currents at the domain boundaries are affected by the normal component of current at that particular boundary. The field observation results by Eigenheer and Quadfasel (2000); Schott and McCreary (2001); Shankar et al. (2002) and model output shows similar pattern of ocean currents, which is a validation for the open boundary condition that is used in this study.

Ocean currents are affected by the surface winds. Surface currents move west during Northeast monsoon (NEM) or DJF season and prominently towards the east during Southwest Monsoon (SWM) or JJAS season in BoB. Fig. 2 shows seasonally averaged model currents that are compared with SODA reanalysis data. Both model and reanalysis data show westward flow in central BoB during NEM; eastward flow in central BoB during SWM. Furthermore, BoB basin exchanges water with AS during SWM and NEM. During NEM water leaves from south-west boundary and during SWM water enters the BoB basin from the same region also seen by Cutler and Swallow (1984); Eigenheer and Quadfasel (2000); Schott and McCreary (2001). The exchange of water mass from BoB to AS and rest of the IO is clearly observed in the model through flow of current near Sri Lanka and equatorial region.

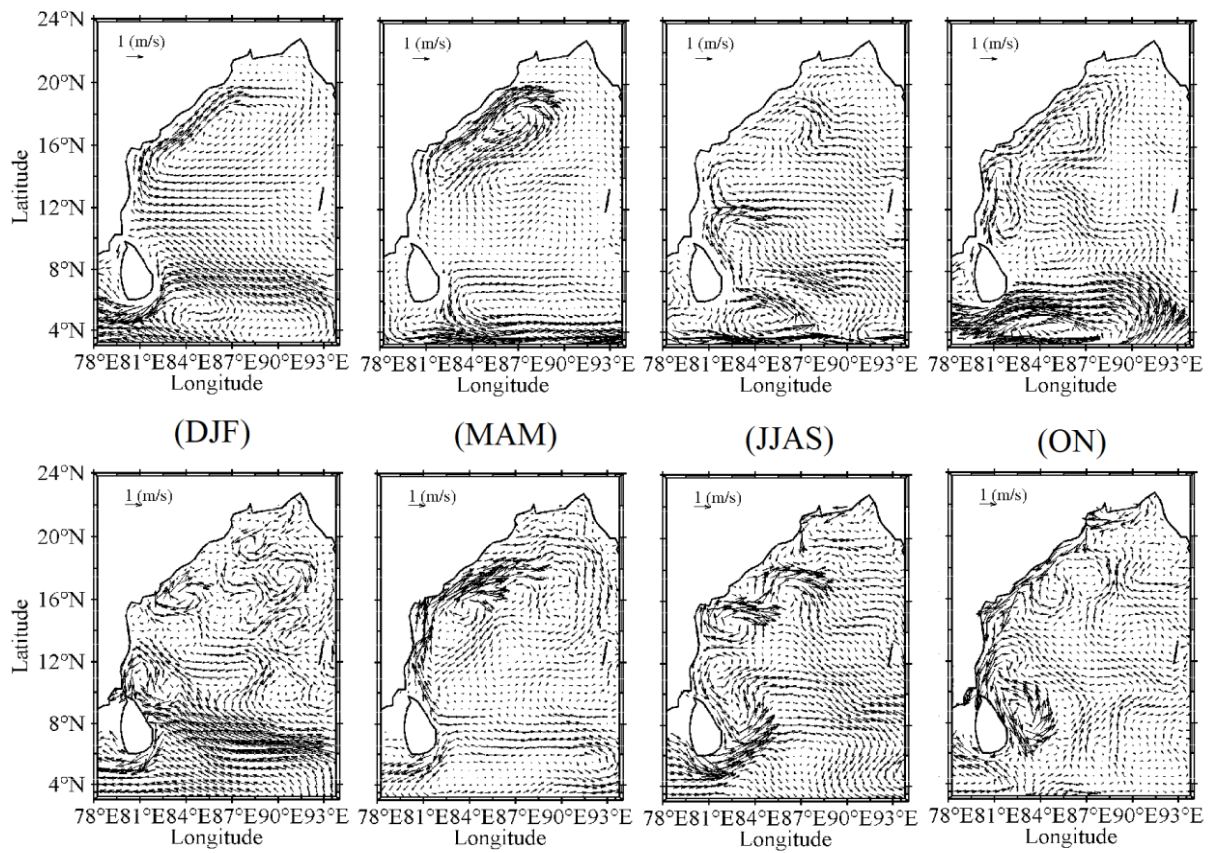

Figure 2: Seasonally averaged surface currents. First row is model output; second row is SODA data.

It is observed that, in BoB, seasonal reversal of wind near the west boundary causes the EICC to reverse its direction. The EICC reverses direction twice a year, flowing north-eastward (pole-ward) from February, with a strong peak in March-April, and south-westward (equator-ward) from October, with the strongest flow in November. This is broadly consistent with the EICC trends reported in Shetye et al. (1991, 1996); Shankar et al. (1996, 2002); Vinayachandran et al. (1996); Schott and McCreary (2001); Behara and Vinayachandran (2016)). Eigenheer and Quadfasel (2000) observed that at the on-set of SWM, flow of EICC weakens and in July, when SWM attains full strength, EICC weakens further. This signifies that the relationship between ocean winds and ocean currents is not simple and direct. From Fig. 2, it is observed that the model is able to reproduce most of the basin currents and boundary currents, both spatially and temporally. There exists a deviation between the magnitudes of model and SODA currents, but the overall directions and trends are very similar. This further gives confidence that radiation condition has been implemented appropriately.

## 3.2 Sea surface temperature and salinity

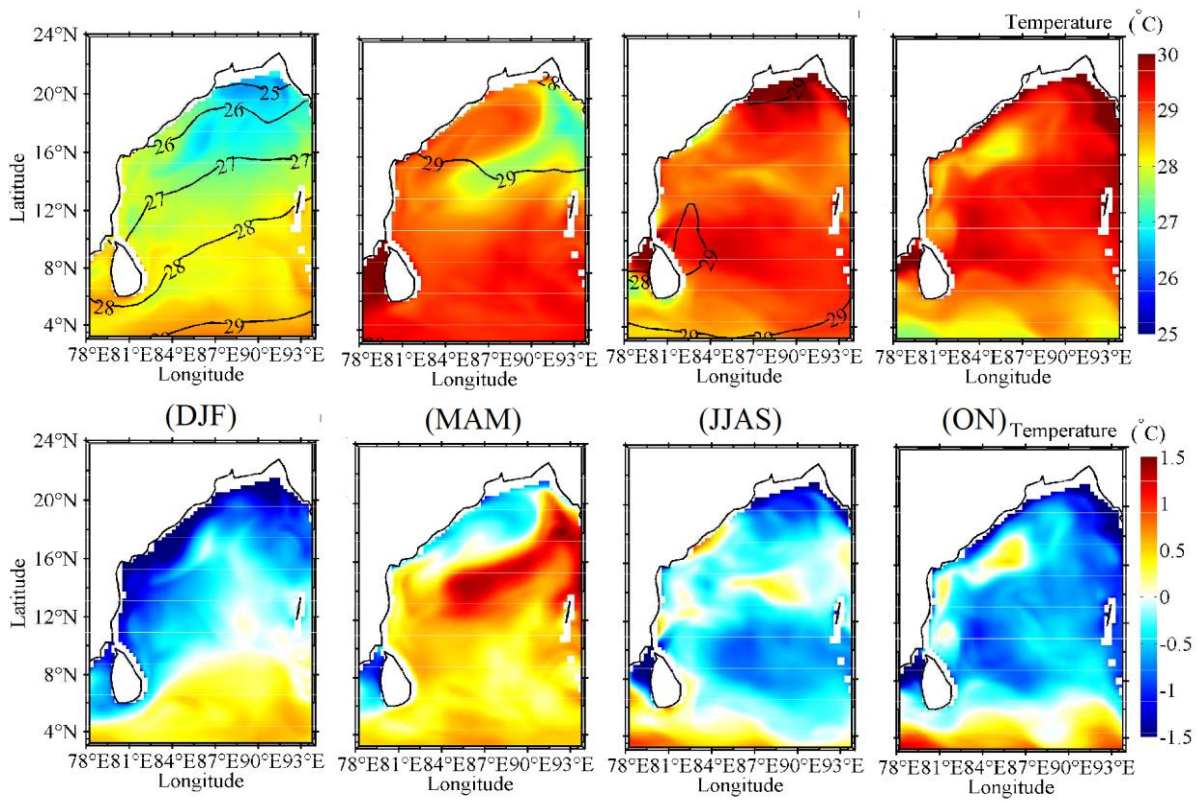

Figure 3: Model SST comparison with SODA reanalysis data. First row shows model output with contour lines of reanalysis data and second row indicate bias (SODA - model).

Sea surface temperature (SST) is one of the important property of ocean for representing ocean stability and precipitation patterns (Uvo et al., 1998;Li et al., 2016). Average SST in BoB is generally on the lower side during winter (ON) and becomes higher during summer (MAM) due to the seasonally varying solar radiation received at the surface. The SST plots shown in Fig. 3 depict higher values of SST (by ~ 1-2 $^{\circ}$C) during summer than post-monsoon (ON), which is also supported by SODA reanalysis data. The warmer temperature at south-eastern part of the bay is seen in Fig. 3 for all seasons from model and SODA dataset. These figures show the SST bias (SODA - model) produced in model simulations. During summer, the entire domain heats up and maintains a temperature of ~ 28 - 30 $^{o}$C due to positive net heat flux. In summer monsoon (JJAS), signature of coastal upwelling is seen with bias of ~ 1 $^{o}$C and it gradually decreases in the post-monsoon (ON) season, a trend also observed by Sarkar et al. (2019); Tirodkar et al. (2020a). In the central domain, the bias ranges between 0 - ±0.5 $^{o}$C throughout the year.

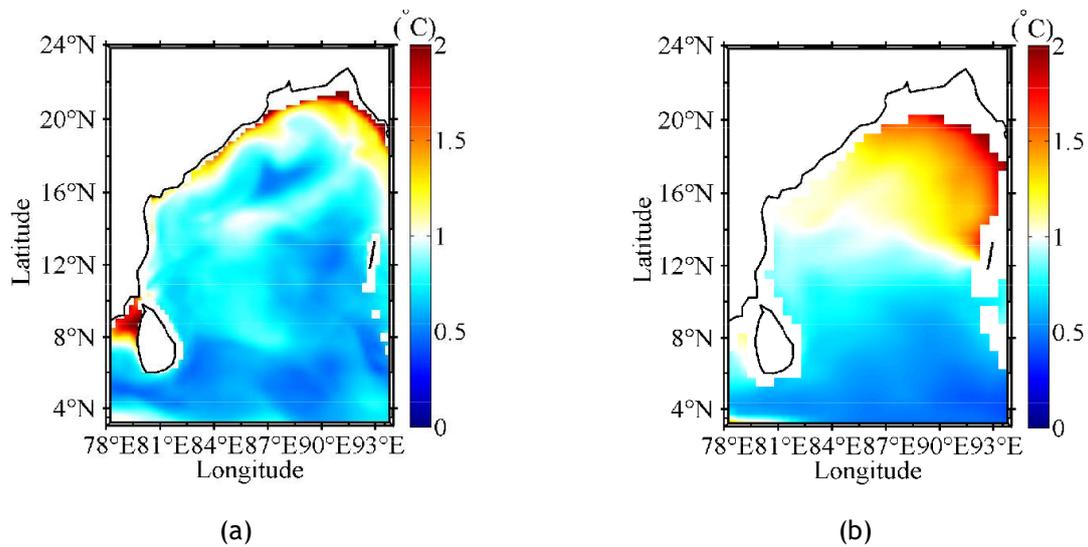

(a)                                  (b)

Figure 4: RMSE in model generated annual SST with respect to (a) SODA data (b) NIOA data

Temperature gradient is observed across latitudes in the bay during north-east monsoon (DJF). This temperature gradient is visible in SODA and NIOA datasets. It occurs due to upwelling at northern coast of bay and the addition of fresh water flux. The root mean square error (RMSE) with NIOA data and SODA data are calculated and shown in Fig. 4. In general, the RMSE values are lower except in the north-eastern part of the bay, which is also documented in Behara and Vinayachandran (2016). This large deviation is due to the absence of river runoff forcing in the model.

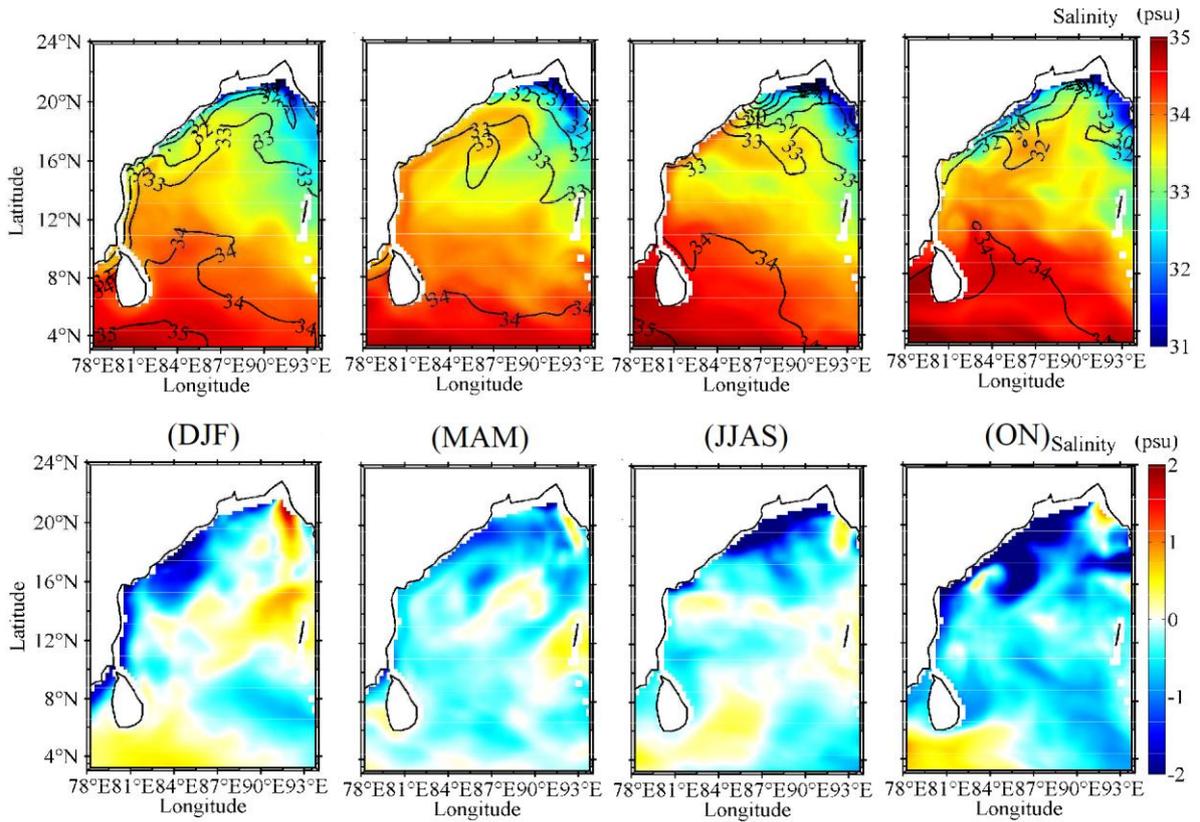

Figure 5: Model SSS comparison with SODA reanalysis data. First row shows model output with contour lines of reanalysis data and second row indicate bias (SODA - model).

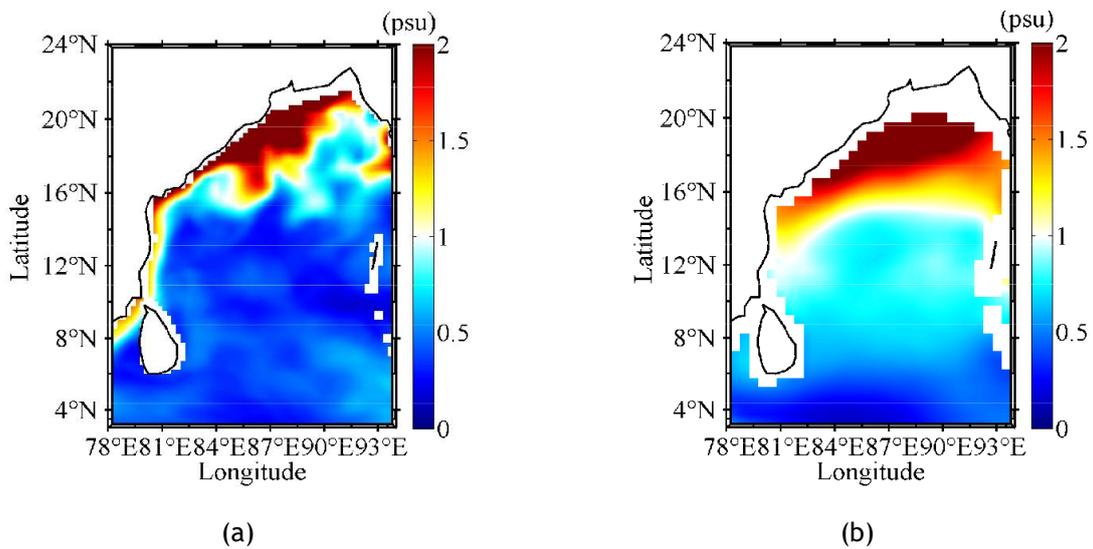

(a)                                      (b)

Figure 6: RMSE in model generated annual SSS with respect to (a) SODA data (b) NIOA data

High amount of river runoff and precipitation (fresh water flux) affects the salinity of the Bay region and therefore, SSS plays an important role in the Bay of Bengal dynamics. Fig. 5 shows model salinity comparison with SODA data. SODA reanalysis considers the effect of fresh water inputs that contribute less saline water in the head bay. Northern and eastern part of BoB gets fresher water input from the rivers like Ganges-Brahmaputra-Mahanadi in north and north-western bay, and Irrawaddy in east (Shetye et al., 1991; Howden and Murtugudde, 2001; Behara and Vinayachandran, 2016). The absence of river runoff in model forcing has lead to biases in model output and is directly visible in SSS pattern. This creates the higher negative SSS bias (reanalysis - model) in the model simulated SSS, especially in the northern part of BoB. These biases are significantly lower in the central bay, indicating satisfactory agreement with the model output and reanalysis data. Fig. 6 shows RMSE of surface salinity with SODA data and NIOA data. The computed salinity is better in rest of the domain, except head bay. The higher RMSE near head bay is due to absence of river runoff forcing.

## 3.3 Vertical structure of temperature and salinity

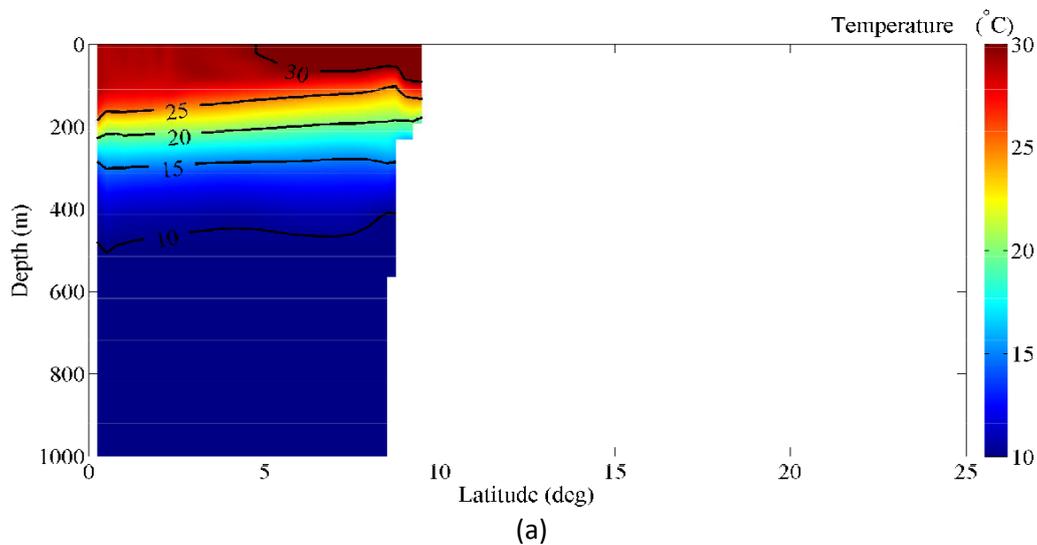

(a)

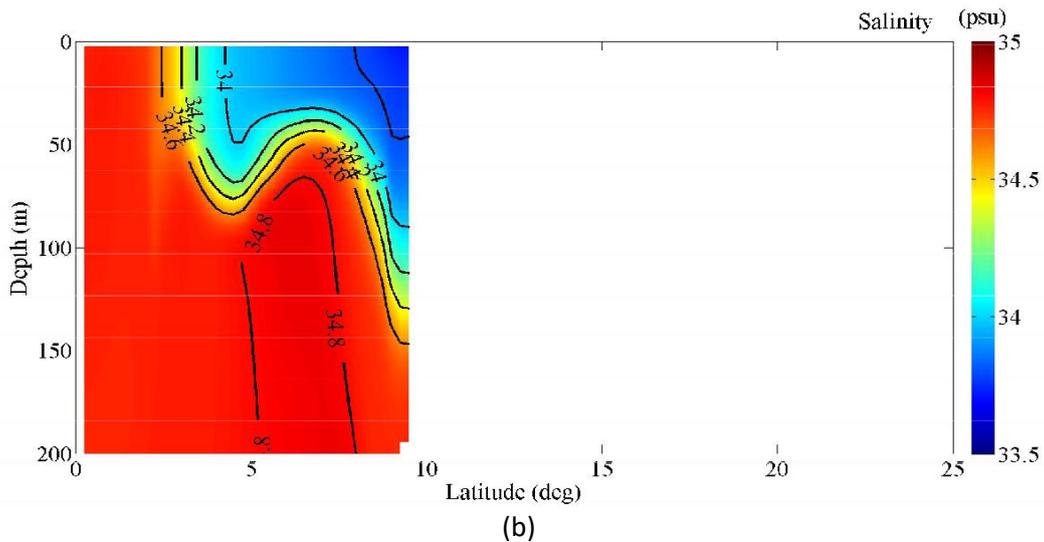

Figure 7: Vertical structure of (a) temperature and (b) salinity at west boundary (75° E) for JJAS season (contour lines only for visualization).

Vertical structures of temperature and salinity helps in understanding the temperature inversions. Figs. 7 and 8 shows the temperature and salinity profiles of ocean at the west boundary and near the equator, respectively. Fig. 7 shows that there is sign of upwelling during south-west monsoon (JJAS) near Indian peninsula and same feature is observed in the seasonal plots of temperature and salinity reanalysis data. These temperature and salinity profiles do not show any spurious pattern with depth. These plots are crucial to understand the deepening of MLD near the equator. Thin mixed layer in northern bay favours temperature inversion by increasing shortwave penetration below the mixed layer. Model output shows temperature inversion in the northern Bay during summer and winter, which is in line with the pattern of temperature inversion observed by Shetye et al. (1996); Thadathil et al. (2002); Vinayachandran et al. (2002).

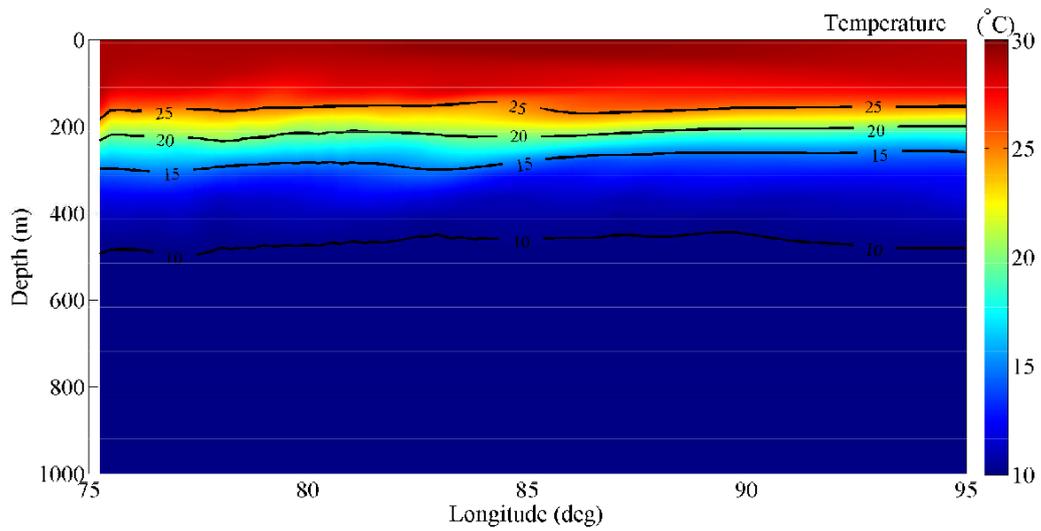

(a)

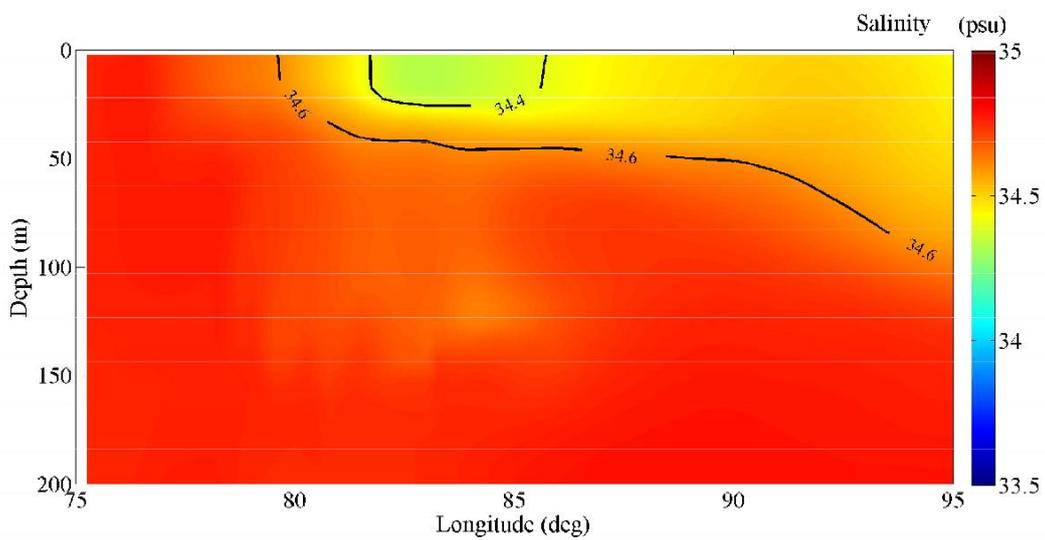

(b)

Figure 8: Vertical structure of (a) temperature and (b) salinity near the equator (1.125° N) for JJAS season (contour lines only for visualization).

## 3.4 Mixed layer depth

The mixed layer depth (MLD) is a measure used to understand the amount of mixing that has occurred in the water column. Here, we use the temperature criteria proposed by de Boyer Montegut et al. (2004) to compute MLD.

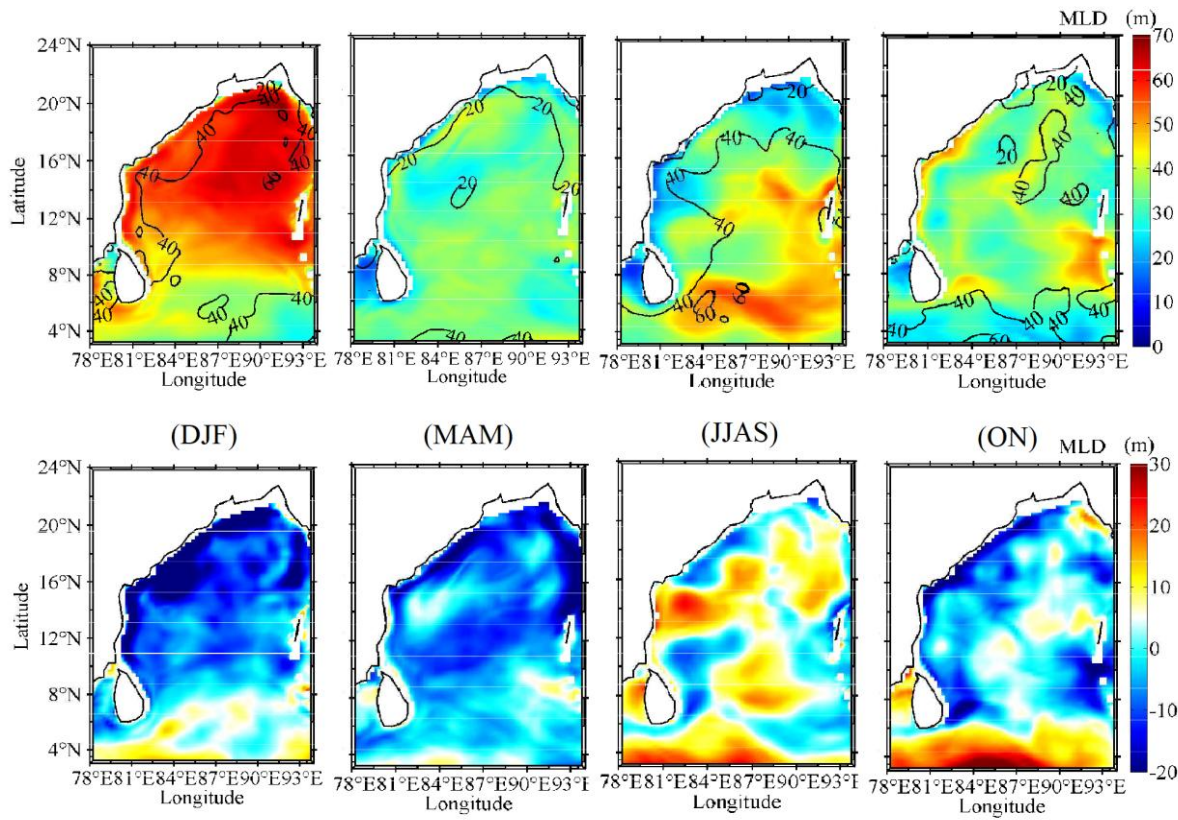

Figure 9: Model MLD comparison with SODA reanalysis data. First row shows model output with contour lines of reanalysis data and second row indicate bias (SODA - model).

Fig. 9 shows model generated MLD, and the bias (SODA - Model) in it. During MAM, model generated MLD is relatively lower. SODA also shows similar seasonality for the MLD patterns. These MLD seasonal patterns show positive correlation with seasonal winds, more the wind stress deeper the MLD. In MAM and ON the bias ranges between ~ 0 - 10 m, whereas in JJAS and DJF the bias is more due to the stronger winds during the monsoon season. In the southern part of the domain, the bias is ~ 10 m - 20 m in all the seasons, except MAM. Summer and winter monsoon model computed MLD is over-estimated by ~ 10 - 30 m compared to SODA MLD showing more mixing in the basin. Despite these differences in the magnitude between model and SODA MLD, the patterns in model MLD are well-represented.

### 3.5 Radiation Heat Budget

Radiation heat budget includes net shortwave radiation, net long wave radiation, sensible heat, and latent heat. In the present model study, the latent heat is computed from evaporation rate. All the heat added to ocean surface layer is considered positive and heat leaving the surface is considered negative. Net shortwave and long wave radiation considered incoming and outgoing radiation balance. Chlorophyll based Morel and an Antoine (1994) shortwave penetration criterion is used to understand the effect of shortwave radiation on the vertical structure of the ocean temperature. All other components of heat budget only affect the ocean surface. Surface heat flux $Q_{net}$ is calculated using the following bulk formula of Louis (1982):

$$Qnet = Qshortwave - Qlongwave - Qsensible - Qlatent \qquad (3)$$

Fig. 10 shows temporal evolution of radiation heat budget at the center of the domain. Both model and WHOI OAFlux data are in good agreement for the entire individual component of radiation heat budget and net heat balance term (Tirodkar et al., 2020b). There is small deviation of shortwave radiation from WHOI OAflux dataset leading to deviation in net heat flux.

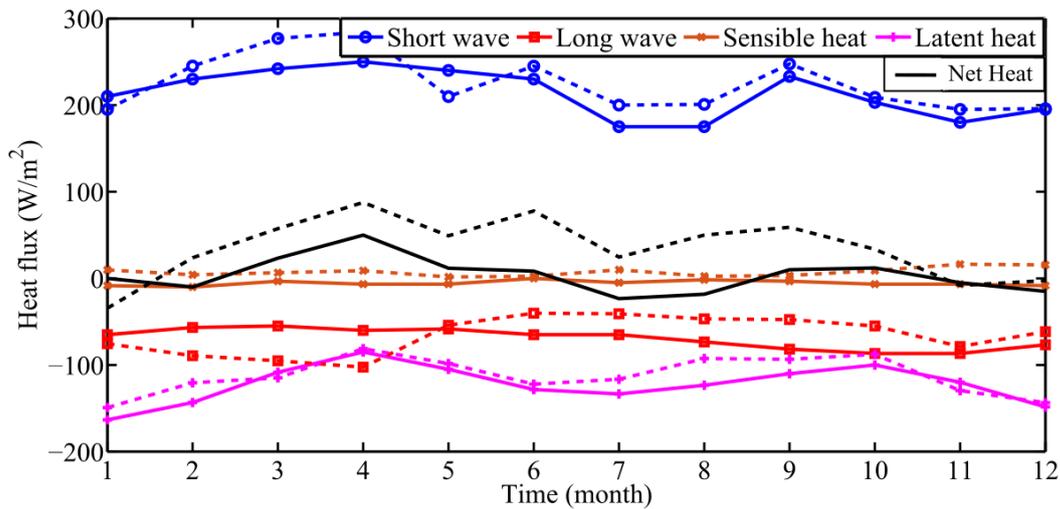

Figure 10: Solar radiation heat budget. Solid line: Model output, Dash line: WHOI data. (Monthly average over 5 years span)

The analysis and comparison of model output like; currents, SST, SSS, vertical profiles of temperature and salinity, MLD, and radiation heat budget, shows that the selected set of parametrization for radiation boundary condition over the domain is suitable for investigation of such small domain. The radiation condition works well with prescription for tracers at boundary. The results simulated are in good agreement with validation datasets.

# 4   SEASONAL VARIABILITY OF ENERGETICS

The flow energetics play an important role in governing the vertical structure of the ocean. The evolution of Turbulent kinetic energy (TKE), formulated below in Eq.4 provides a mechanistic approach to understand the role of turbulent flux on ocean processes and vertical stability (Pope, 2000).

$$\underbrace{\frac{\partial K}{\partial t} + \overline{U}_J \frac{\partial K}{\partial x_j}}_{I} = \underbrace{-\frac{1}{\rho_0}\frac{\partial \overline{u'_j u'_j u'_i}}{\partial x_i} + \nu \frac{\partial^2 K}{\partial x_j^2}}_{II} \underbrace{- \overline{u'_i u'_j} \frac{\partial \overline{u_i}}{\partial x_j}}_{III} \underbrace{- \frac{g}{\rho_0} \overline{\rho' u'_i} \delta_{i3}}_{IV} \underbrace{- \nu \overline{\frac{\partial u'_i}{\partial x_j}\frac{\partial u'_i}{\partial x_j}}}_{V} \qquad (4)$$

where, $\overline{U}$ is the mean flow velocity, u', p' and ρ' are the fluctuating components of velocity, pressure, and density, u is the instantaneous velocity, $\rho_0$ is the reference density (taken at the surface), ν is the kinematic viscosity and $K = \frac{1}{2}(\overline{u'^2} + \overline{v'^2} + \overline{w'^2})$ is the turbulent kinetic energy.

In Eq.4, term (I) represents the temporal and spatial gradients of K, term (II) accounts for the transport of K from one region to another, term (III) is the production of energy ($P = -\overline{u'_i u'_j}\frac{\partial \overline{u_i}}{\partial x_j}$), which is the primary source term due to wind shear, term (IV) is the buoyancy flux ($B = \frac{g}{\rho_0}\overline{\rho' u'_i}\delta_{i3}$), which is source of potential energy coming from the stratification, and term (V) is the viscous dissipation ($\varepsilon = \nu \overline{\frac{\partial u'_i}{\partial x_j}\frac{\partial u'_i}{\partial x_j}}$), which is the sink term accounting for energy dissipating in the flow. It is well accepted that the flow energetics could be understood by looking at P, B, ε as they govern the source and sink for TKE. All other terms account only for the advection of TKE, which is secondary for analysis of energetics. In this section the analysis of spatial plots of fluxes, namely, K(x,y), P(x,y), B(x,y) and ε(x,y) are discussed to understand effect of seasonal wind pattern on the overall energy budget and how it affects the vertical mixing and structure.

The following methodology is employed to get the seasonal plots of TKE budget terms. First, the velocity and density fields are time averaged over the particular season (say JJAS) to obtain a single mean field. Any mean field, say $\overline{A}$, takes the form $\overline{A}(x,y,z) = \frac{1}{T}\int adt$, where T is the total duration and dt is the time step at which each field is available. The fluctuation field is obtained as $a'(x,y,z,t) = a(x,y,z,t) - \overline{A}(x,y,z)$. The fluctuation fields are then used for calculation of K, P, B, and ε. Finally, these quantities are time-averaged and 30 m depth averaged to obtain the seasonal plots of the energetics.

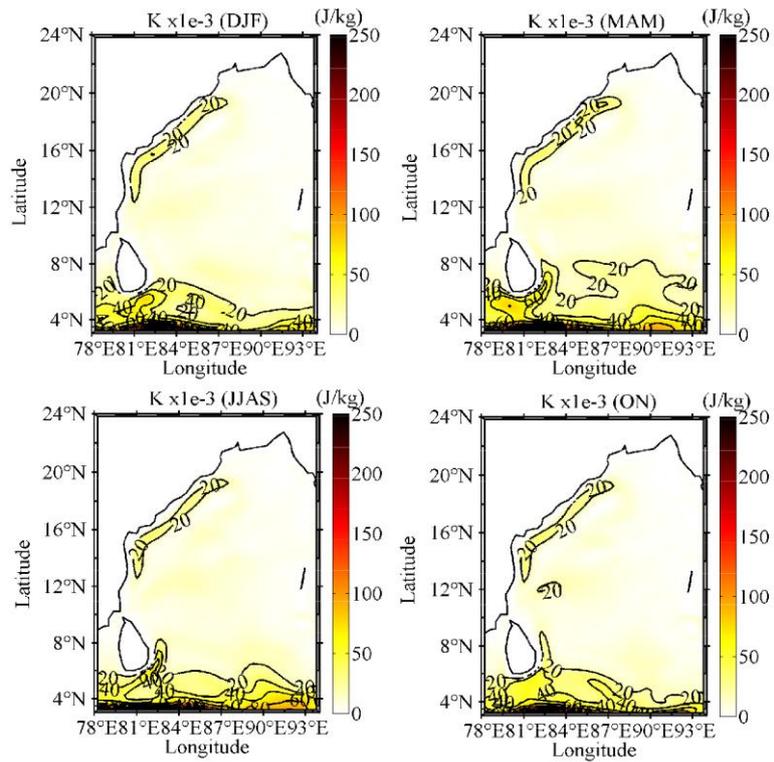

Figure 11: Seasonal variation of turbulent kinetic energy (K). All plots are averaged over the respective season and depth.

Figure 11 shows the seasonal average of K over Bay of Bengal. In the southern boundary, owing to strong currents, the energy is high during all the seasons, which is indicative of more mixing in this region. During the pre-monsoon (MAM) and post-monsoon (ON), the presence of EICC provides higher energy along the eastern coast of India.

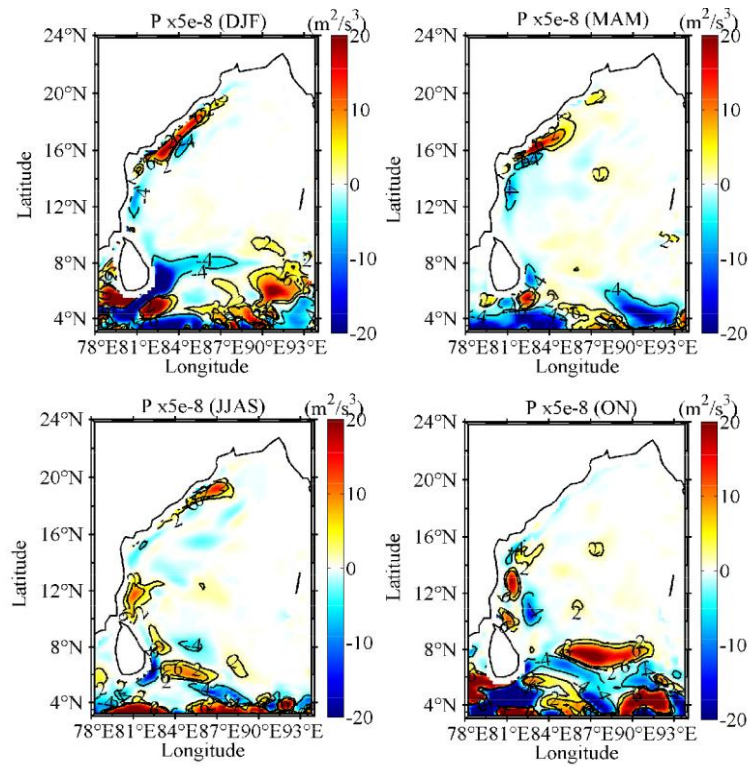

Figure 12: Seasonal variation of production flux (P). All plots are averaged over the respective season and depth.

In order to delve into the energetics, the flux terms in the TKE budget equation need to be individually diagnosed to get a better understanding of the process dynamics. First we will start with the production flux, P, which presents the mechanical or shear production or loss. The production flux term usually signifies the transfer of energy from the mean flow to the turbulence, i.e. cascade of energy from large-scales to small-scales. This forward energy cascade occurs when P is positive, meaning it aids in production of turbulence. However, there are situations when P becomes negative, in which case the energy is extracted from small-scale to large-scale; also known as "inverse energy-cascade" (Tennekes and Lumley, 2018). When P is negative, the energy is transferred from the turbulence to the mean flow, which means the mean flow energizes and maintains its structure. Now, in Figure 12, we see that P is largely non-existent (i.e. very close to zero) in the Central BoB. Near the east coast, P is positive, which is again due to the presence of EICC. The interesting behavior in P is observed near the southern boundary, where large negative values are seen along with some positive values. This indicates that near the southern boundary, owing to the inverse energy-cascade effect, the mean flow energizes that allow the meso-scale and synoptic-scale structures to extract energy from the turbulence. This dynamically means that the mean flow structures, such as, large-scale circulation patterns have a tendency to retain their structure near the southern boundary compared to other parts of BoB. This analysis leads us to believe that southern BoB possibly is the region where large-scale (mean flow) eddies persist and transport energy to other parts of the Bay. The advection of energy from these large-scale eddies would play an important role in controlling the vertical and horizontal mixing in other regions of BoB.

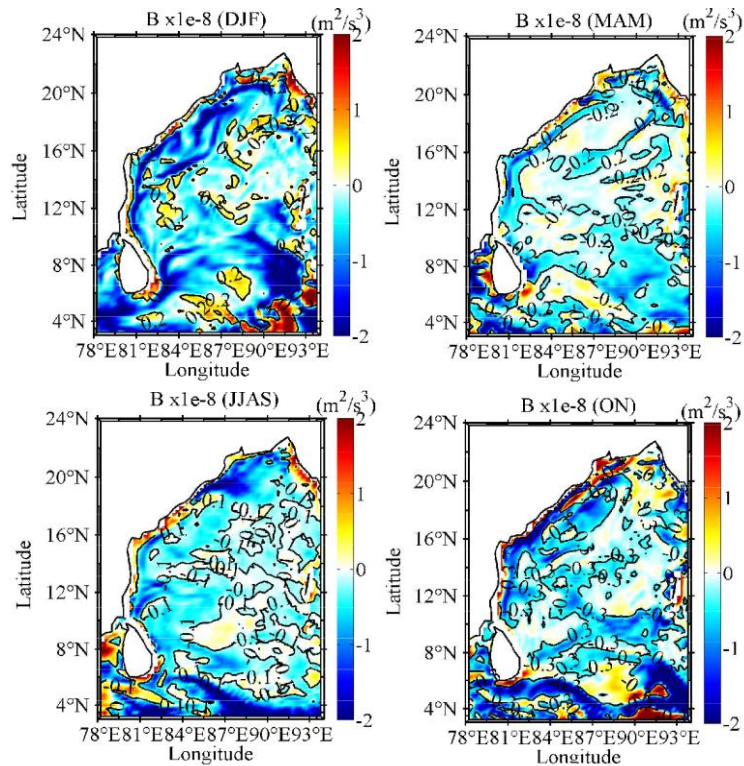

Figure 13: Seasonal variation of buoyancy flux (B). All plots are averaged over the respective season and depth.

The buoyancy flux term, B is a representation of the potential energy stored in that region. Positive values of B represent stable stratification (high potential energy state) and negative values of B mean unstable stratification (low potential energy state). Looking at the buoyancy flux plots in Figure 13, it is clearly visible that B is predominantly negative in BoB during all seasons. This is indicative of scalar driven mixing due to unstable stratification, which again aids in the production of turbulence. This also shows that B is the dominant source of turbulence in the North and Central BoB regions, where P was almost zero. Our analysis shows that the mixing and

MLD in head Bay and Central Bay is governed by the buoyancy fluxes and presence of unstable stratification in these regions.

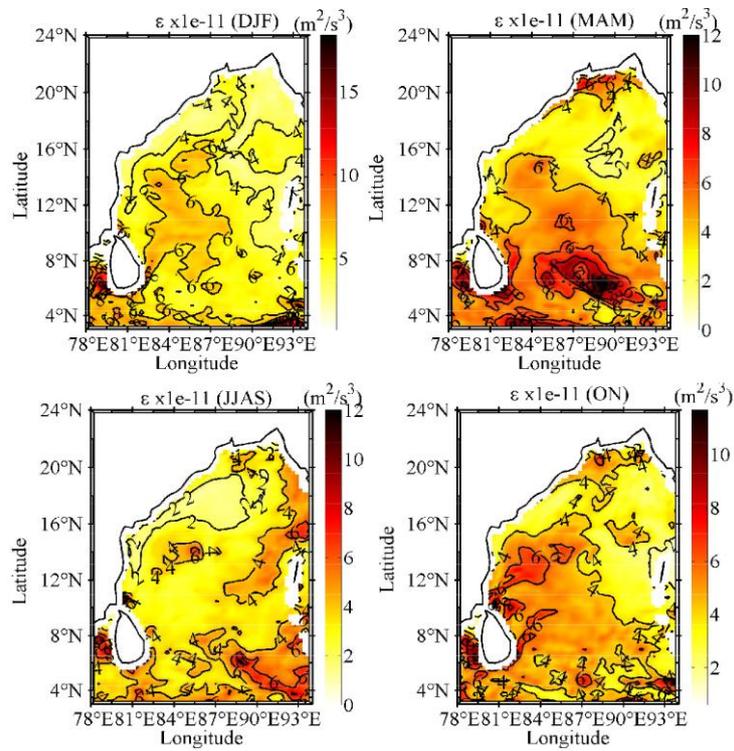

Figure 14: Seasonal variation of viscous dissipation ($\varepsilon$). All plots are averaged over the respective season and depth.

Finally, we look at the viscous dissipation, $\varepsilon$, which is defined as the irreversible process by which energy from small-scales are dissipated to heat, putting an end to the turbulence process. Turbulence source terms such as P and B feed to $\varepsilon$, which is the sink of energy. Figure 14 shows that $\varepsilon$ is high in the southern BoB, where the K is also high. During pre-monsoon (MAM) and post-monsoon (ON) seasons, $\varepsilon$ shows high values near the east coast, which is due to the EICC presence, which enhances turbulence activity and hence $\varepsilon$.

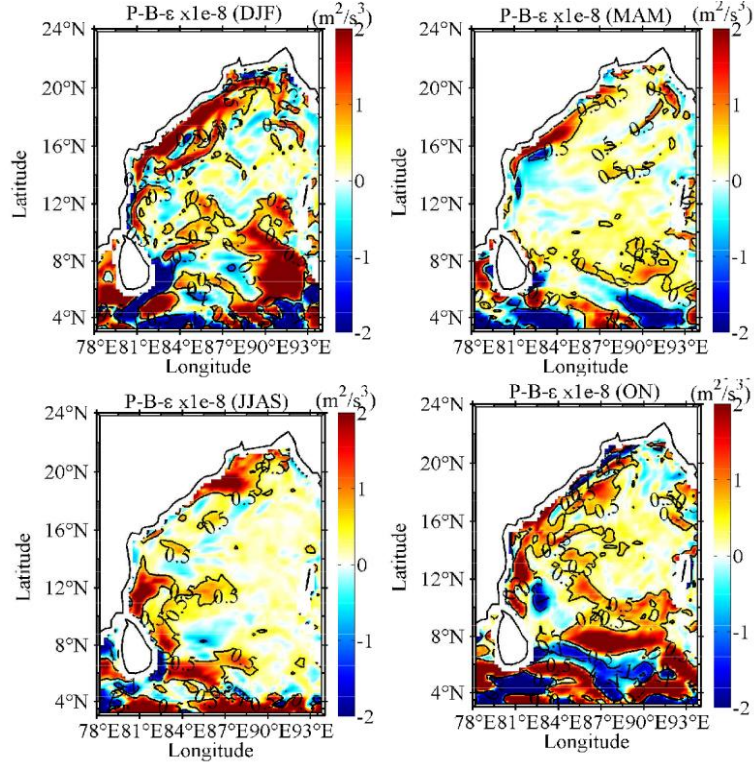

Figure 15: Seasonal variation of P-B-ε.

Rewriting Eq. 4 assuming stationary conditions (i.e. $\frac{\partial}{\partial t} + \overline{U}_j \frac{\partial}{\partial x_j} = 0$), we get,

P-B-ε- $D_K$ = 0  (5)

where $D_K$ accounts for all the transport terms (term II in Eq. 4). If transport is zero, then we would get a balance between the other flux terms, namely, P-B-ε =0. Plotting this term in Figure 15, we see a non-zero transport term during all the seasons. This figure further confirms the fact that the transport/advection of energy occurs from southern BoB region, where predominantly positive $D_K$ values are observed. During pre-monsoon seasons, the transport of energy due to EICC is also seen. The transport term is low over central and north Bay, which indicate that these are the regions of stationary and homogeneous turbulence, predominantly governed by a balance of turbulent fluxes, namely, P, B and ε. The results from the energetics analysis clearly show regional variations in process-driving mixing dynamics in BoB, which is important to come up with region specific parametrizations for mixing and turbulence.

# 5 Conclusion

Modular Ocean Model (MOM) with open boundary condition (OBC) for modeling the regional dynamics in the Bay of Bengal domain forced with air-sea fluxes. The domain, under consideration, was initialized with temperature and salinity profiles and was allowed to spin-up for 10 years with climatological forcing. Following this, 5-year model outputs were simulated and the results were compared with SODA reanalysis data. The results show successful implementation of open boundary condition in the domain with assigned parametrizations. Conservation of mass was achieved inside the domain with a stable profile of global averaged kinetic energy. The smooth exchange of tracers across boundary was observed with the help of vertical temperature and salinity profiles.

The seasonal variability due to reversal of winds and solar forcing was observed from spatial and temporal plots of SST, SSS, and MLD. We clearly observed that in the absence of river runoff forcing, near the head bay, biases in temperature and salinity were high compared to other sections of the Bay. Model comparison with validation data sets were overall in good agreement, which was also observed from the RMSE plots of SST and SSS. Mixing characteristics of model parameters were studied with the help of MLD. It showed that the model was able to simulate seasonal changes in ocean stratification and its vertical structure. Radiation heat budget showed that seasonal effect of air-sea flux was successfully captured by the model, which was validated with the WHOI-OA flux dataset. The above results reveal that open boundary condition worked well in the regional domain under consideration.

After validation of OBC for a regional domain in the north Indian Ocean, the ensuing flow energetics based on the turbulent flux terms was carried out. The analysis clearly indicated the presence of inverse energy-cascade in the southern Bay of Bengal, wherein energy flows back into the mean flow structures (such as large-scale eddies or circulation currents). Further analysis shows that the B term is the major source of turbulence production in the north and central Bay of Bengal regions. These results convey mechanisms by which energy is produced and transferred in different regions of Bay of Bengal, which has important bearing in modeling of Indian Summer Monsoon.

Our study claims that, OBC has been implemented for the Bay of Bengal basin for the first time using MOM5. This work open up the possibility to perform sub-mesoscale resolution runs in small regional domains using open boundary conditions rather than simulating larger domain with damping zone near boundary (e.g. sponge condition), which are computationally costly and time-consuming. The energetics study also reveals that OBC would help in capturing the effect of complex ocean processes, whose efficacy has been proved in this work.

# 6 Acknowledgements

The authors acknowledge funding from ESSO - Indian National Centre for Ocean Information Services, India and help from project staff Mr. V L Srinivas who was supported through this funding. We also acknowledge support from Department of Science and Technology under NMSKCC scheme for this work.

# Reference

Behara, A., Vinayachandran, P.N., 2016. An ogcm study of the impact of rain and river water forcing on the bay of bengal. Journal of Geophysical Research: Oceans 121, 2425{2446. URL: http://dx.doi.org/10.1002/2015JC011325, doi:10.1002/2015JC011325.

Bhat, G., 2003. Some salient features of the atmosphere observed over the north bay of bengal during bobmex. Journal of Earth System Science 112,131-146.

Bhat, G., Gadgil, S., Hareesh Kumar, P., Kalsi, S., Madhusoodanan, P., Murty, V., Prasada Rao, C., Babu, V.R., Rao, L., Rao, R., et al., 2001.Bobmex: The bay of bengal monsoon experiment. Bulletin of the American Meteorological Society 82, 2217-2244.

de Boyer Mont egut, C., Madec, G., Fischer, A.S., Lazar, A., Iudicone, D., 2004. Mixed layer depth over the global ocean: An examination of profile data and a profile-based climatology. Journal of Geophysical Research: Oceans 109.

Bryan, K., Cox, M.D., 1967. A numerical investigation of the oceanic general circulation. Tellus 19, 54-80.

Bryan, K., Cox, M.D., 1972. An approximate equation of state for numerical models of ocean circulation. Journal of Physical Oceanography 2, 510-514.

Carton, J.A., Chepurin, G., Cao, X., Giese, B., 2000. A simple ocean data assimilation analysis of the global upper ocean 1950-95. part i: Methodology. Journal of Physical Oceanography 30, 294-309.

Carton, J.A., Chepurin, G.A., Chen, L., 2018. SODA3: A new ocean climate reanalysis. Journal of Climate 31, 6967-6983.

Chai, T., Draxler, R.R., 2014. Root mean square error (rmse) or mean absolute error (mae)?-arguments against avoiding rmse in the literature. Geoscientific Model Development Discussions 7, 1247-1250.

Chassignet, E.P., Garraffo, Z.D., 2001. Viscosity parameterization and the Gulf Stream separation. Technical Report. Miami Univ Fl Inst Of Marine And Atmospheric Sciences.

Chatterjee, A., Shankar, D., McCreary, J., Vinayachandran, P., Mukherjee, A., 2017. Dynamics of andaman s ea circulation and its role in connecting the equatorial Indian Ocean to the bay of bengal. Journal of Geophysical Research: Oceans 122, 3200-3218.

Chatterjee, A., Shankar, D., McCreary Jr, J., Vinayachandran, P., 2013. Yanai waves in the western equatorial indian ocean. Journal of Geophysical Research: Oceans 118, 1556-1570.

Chatterjee, A., Shankar, D., Shenoi, S., Reddy, G., Michael, G., Ravichandran, M., Gopalkrishna, V., Rao, E.R., Bhaskar, T.U., Sanjeevan, V., 2012. A new atlas of temperature and salinity for the north indian ocean. Journal of Earth System Science 121, 559-593.

Cheng, X., McCreary, J.P., Qiu, B., Qi, Y., Du, Y., Chen, X., 2018. Dynamics of eddy generation in the central bay of bengal. Journal of Geophysical Research: Oceans 123, 6861-6875.

Courtois, P., Hu, X., Pennelly, C., Spence, P., Myers, P.G., 2017. Mixed layer depth calculation in deep convection regions in ocean numerical models. Ocean Modelling 120, 60-78.

Cox, M.D., Bryan, K., 1984. A numerical model of the ventilated thermocline. Journal of Physical Oceanography 14, 674-687.

Cui, W., Yang, J., Ma, Y., 2016. A statistical analysis of mesoscale eddies in the bay of bengal from 22-year altimetry data. Acta Oceanologica Sinica 35, 16-27.

Cutler, A., Swallow, S., 1984. Surface currents of the indian ocean (to 25 s, 100 e) compiled from historical data archived by the uk meteorological office, rep. 187, 36 pp. Inst. of Oceanogr. Sci., Godalming, UK .

Effy, J.B., Francis, P., Ramakrishna, S., Mukherjee, A., 2020. Anomalous warming of the western equatorial indian ocean in 2007: Role of ocean dynamics. Ocean Modelling 147, 101542.

Eigenheer, A., Quadfasel, D., 2000. Seasonal variability of the Bay of Bengal circulation inferred from TOPEX/Poseidon altimetry. Journal of Geophysical Research: Oceans 105, 3243-3252.

Francis, P., Vinayachandran, P., Shenoi, S., 2013. The indian ocean forecast system. Current Science(Bangalore) 104, 1354-1368.

Francis, P.A., Jithin, A.K., Chatterjee, A., Mukherjee, A., Shankar, D., Vinayachandran, P.N., Ramakrishna, S.S.V.S., 2020. Structure and dynamics of undercurrents in the western boundary current of the bay of bengal. Ocean Dynamics 70, 387-404.

Fu, L.L., Christensen, E.J., Yamarone, C.A., Lefebvre, M., Menard, Y., Dorrer, M., Escudier, P., 1994. TOPEX/Poseidon mission overview. Journal of Geophysical Research: Oceans 99, 24369-24381.

Griffies, S., 2003. Fundamentals of ocean climate models. Princeton university press.

Griffies, S.M., 2012. Elements of the modular ocean model (mom). GFDL Ocean Group Tech. Rep 7, 620.

Griffies, S.M., Biastoch, A., Bo•ning, C., Bryan, F., Danabasoglu, G., Chassignet, E.P., England, M.H., Gerdes, R., Haak, H., Hallberg, R.W., et al., 2009. Coordinated ocean-ice reference experiments (cores). Ocean modelling 26, 1-46.

Gulakaram, V.S., Vissa, N.K., Bhaskaran, P.K., 2018. Role of mesoscale eddies on atmospheric convection during summer monsoon season over the bay of bengal: A case study. Journal of Ocean Engineering and Science 3, 343-354.

Gulakaram, V.S., Vissa, N.K., Bhaskaran, P.K., 2020. Characteristics and vertical structure of oceanic mesoscale eddies in the bay of bengal. Dynamics of Atmospheres and Oceans 89, 101131.

Herzfeld, M., Schmidt, M., Griffies, S., Liang, Z., 2011. Realistic test cases for limited area ocean modelling. Ocean Modelling 37, 1-34.

Howden, S.D., Murtugudde, R., 2001. Effects of river inputs into the bay of bengal. Journal of Geophysical Research: Oceans 106, 19825-19843. URL: http://dx.doi.org/10.1029/2000JC000656, doi:10.1029/2000JC000656.

Hu, C., Lee, Z., Franz, B., 2012. Chlorophyll aalgorithms for oligotrophic oceans: A novel approach based on three-band reflectance difference. Journal of Geophysical Research: Oceans 117.

Jensen, T.G., 1998. Open boundary conditions in stratified ocean models. Journal of Marine Systems 16, 297-322.

Kundu, P., Cohen, L., 1990. Fluid mechanics, 638 pp. Academic, Calif .

Kurian, J., Vinayachandran, P., 2007. Mechanisms of formation of the arabian sea mini warm pool in a high-resolution ocean general circulation model. Journal of Geophysical Research: Oceans 112.

Kurien, P., Ikeda, M., Valsala, V.K., 2010. Mesoscale variability along the east coast of india in spring as revealed from satellite data and ogcm simulations. Journal of oceanography 66, 273-289.

Large, W.G., McWilliams, J.C., Doney, S.C., 1994. Oceanic vertical mixing: A review and a model with a nonlocal boundary layer parameterization. Reviews of Geophysics 32, 363-403.

Large, W.G., Yeager, S.G., 2004. Diurnal to decadal global forcing for ocean and sea-ice models: The data sets and flux climatologies.

Li, Y., Han, W., Wang, W., Ravichandran, M., 2016. Intraseasonal variability of sst and precipitation in the arabian sea during the indian summer monsoon: Impact of ocean mixed layer depth. Journal of Climate 29, 7889-7910.

Locarnini, M., Mishonov, A., Baranova, O., Boyer, T., Zweng, M., Garcia, H., Seidov, D., Weathers, K., Paver, C., Smolyar, I., et al., 2018. World ocean atlas 2018, volume 1: Temperature .

Louis, J., 1982. A short history of the operational pbl-parameterization at ecmwf, in: ECMWF Workshop Planetary Boundary Layer Parameterization, 1982, ECMWF.

Marsaleix, P., Auclair, F., Estournel, C., 2006. Considerations on open boundary conditions for regional and coastal ocean models. Journal of Atmospheric and Oceanic Technology 23, 1604-1613.

Masumoto, Y., Meyers, G., 1998. Forced rossby waves in the southern tropical indian ocean. Journal of Geophysical Research: Oceans 103,593 27589-27602. URL: http://dx.doi.org/10.1029/98JC02546, doi:10.1029/98JC02546.

McCreary, J.P., Kundu, P.K., Molinari, R.L., 1993. A numerical investigation of dynamics, thermodynamics and mixed layer processes in the indian ocean. Progress in Oceanography 31, 181-244. URL: http://www.sciencedirect.com/science/article/pii/007966119390002U, doi:http://dx.doi.org/10.1016/0079-6611(93)90002-U.

Morel, A., Antoine, D., 1994. Heating rate within the upper ocean in relation to its bio{optical state. Journal of Physical Oceanography 24, 1652-1665.

Mukherjee, A., Shankar, D., Chatterjee, A., Vinayachandran, P., 2018. Numerical simulation of the observed near-surface east india coastal current on the continental slope. Climate Dynamics 50, 3949-3980.

Nuncio, M., Kumar, S.P., 2012. Life cycle of eddies along the western boundary of the bay of bengal and their implications. Journal of Marine Systems 94, 9-17.

Orlanski, I., 1976. A simple boundary condition for unbounded hyperbolic flows. Journal of Computational Physics 21, 251-269.

Perigaud, C., Delecluse, P., 1992. Annual sea level variations in the southern tropical indian ocean from geosat and shallow-water simulations. Journal of Geophysical Research: Oceans 97, 20169-20178. URL: http://dx.doi.org/10.1029/92JC01961, doi:10.1029/92JC01961.

Pope, S.B., 2000. Turbulent Flows. Cambridge University Press. doi:10.1017/CBO9780511840531.

Potemra, J.T., Luther, M.E., O'Brien, J.J., 1991. The seasonal circulation of the upper ocean in the bay of bengal. Journal of Geophysical Research: Oceans 96, 12667-12683. URL: http://dx.doi.org/10.1029/91JC01045,doi:10.1029/91JC01045.

Prasanna Kumar, S., Nuncio, M., Narvekar, J., Kumar, A., Sardesai, d.S., De Souza, S., Gauns, M., Ramaiah, N., Madhupratap, M., 2004. Are eddies nature's trigger to enhance biological productivity in the bay of bengal? Geophysical Research Letters 31.

Rahaman, H., Venugopal, T., Penny, S.G., Behringer, D.W., Ravichandran, M., Raju, J., Srinivasu, U., Sengupta, D., 2019. Improved ocean analysis for the indian ocean. Journal of Operational Oceanography 12, 16-33.

Roman-Stork, H.L., Subrahmanyam, B., Trott, C.B., 2019. Mesoscale eddy variability and its linkage to deep convection over the bay of bengal using satellite altimetric observations. Advances in Space Research .

Sarkar, M., Chauhan, R.O., Tirodkar, S.A., Balasubramanian, S., Behera, M.R., 2019. Bay of bengal mixing layer dynamics under different seasonal wind scenarios using modular ocean model. AGUFM 2019, OS11D-1512.

Schott, F.A., McCreary, J.P., 2001. The monsoon circulation of the indian ocean. Progress in Oceanography 51, 1-123. URL: http://www.sciencedirect.com/science/article/pii/S0079661101000830, doi:http://dx.doi.org/10.1016/S0079-6611(01)00083-0.

Shaji, C., Iizuka, S., Matsuura, T., 2003. Seasonal variability of near-surface heat budget of selected oceanic areas in the north tropical indian ocean. Journal of oceanography 59, 87-103.

Shankar, D., McCreary, J.P., Han, W., Shetye, S.R., 1996. Dynamics of the east india coastal current: 1. analytic solutions forced by interior ekman pumping and local alongshore winds. Journal of Geophysical Research: Oceans 101, 13975{13991. URL: http://dx.doi.org/10.1029/96JC00559, doi:10.1029/96JC00559.

Shankar, D., Vinayachandran, P., Unnikrishnan, A., 2002. The monsoon currents in the north indian ocean. Progress in Oceanography 52, 63-120. URL: http://www.sciencedirect.com/science/article/pii/S0079661102000241, doi:http://dx.doi.org/10.1016/S0079-6611(02)00024-1.

Shetye, S., Shenoi, S., Gouveia, A., Michael, G., Sundar, D., Nampoothiri, G., 1991. Wind-driven coastal upwelling along the western boundary of the bay of bengal during the southwest monsoon. Continental Shelf Research 11, 1397-1408. URL: http://www.sciencedirect.com/science/article/pii/0278434391900425, doi:http://dx.doi.org/10.656 1016/0278-4343(91)90042-5.

Shetye, S.R., Gouveia, A.D., Shankar, D., Shenoi, S.S.C., Vinayachandran, P.N., Sundar, D., Michael, G.S., Nampoothiri, G., 1996. Hydrography and circulation in the western bay of bengal during the northeast monsoon. Journal of Geophysical Research: Oceans 101, 14011-14025. URL: http://dx.doi.org/10.1029/95JC03307, doi:10.1029/95JC03307.

Sindhu, B., Suresh, I., Unnikrishnan, A.S., Bhatkar, N.V., Neetu, S., Michael, G.S., 2007. Improved bathymetric datasets for the shallow water regions in the Indian Ocean. Journal of Earth System Science 116, 261-274.

Srivastava, A., Dwivedi, S., Mishra, A.K., 2018. Investigating the role of air-sea forcing on the variability of hydrography, circulation, and mixed layer depth in the arabian sea and bay of bengal. Oceanologia 60, 169-186.

Suneet, D., Kumar, M.A., Atul, S., 2019. Upper ocean high resolution regional modeling of the arabian sea and bay of bengal. Acta Oceanologica Sinica 38, 32-50.

Tennekes, H., Lumley, J.L., 2018. A first course in turbulence. MIT press.

Thadathil, P., Gopalakrishna, V., Muraleedharan, P., Reddy, G., Araligidad, N., Shenoy, S., 2002. Surface layer temperature inversion in the bay of bengal. Deep Sea Research Part I: Oceanographic Research Papers 49, 1801-1818.

Thompson, B., Gnanaseelan, C., Salvekar, P., 2006. Variability in the indian ocean circulation and salinity and its impact on sst anomalies during dipole events. Journal of Marine Research 64, 853-880.

Tirodkar, S., Behera, M.R., Balasubramanian, S., 2020a. A regional study of bay of bengal processes using radiation boundary condition in modular ocean model, in: EGU General Assembly Conference Abstracts, p. 513.

Tirodkar, S., Behera, M.R., Balasubramanian, S., 2020b. Regional study of turbulent and radiation fluxes in bay of bengal (bob) using modular ocean model (mom), in: Abstract(A062-0021)AGU Fall Meeting 2020, AGU.

Tsai, P.T.H., O'Brien, J.J., Luther, M.E., 1992. The 26-day oscillation observed in the satellite sea surface temperature measurements in the equatorial western indian ocean. Journal of Geophysical Research: Oceans 97, 9605-9618. URL: http://dx.doi.org/10.1029/91JC03162, 689 doi:10.1029/91JC03162.

Uvo, C.B., Repelli, C.A., Zebiak, S.E., Kushnir, Y., 1998. The relationships between tropical pacific and atlantic sst and northeast brazil monthly precipitation. Journal of Climate 11, 551-562.

Vinayachandran, P., Murty, V., Ramesh Babu, V., 2002. Observations of barrier layer formation in the bay of bengal during summer monsoon. Journal of Geophysical Research: Oceans 107, SRF-19.

Vinayachandran, P., Shetye, S.R., Sengupta, D., Gadgil, S., 1996. Forcing mechanisms of the bay of bengal. Curr. Sci 70, 753-763.

Yu, L., Jin, X., Weller, R., 2008. Multidecade global flux datasets from the Objectively Analyzed Air-Sea Fluxes (OAFlux) project: Latent and sensible heat fluxes, ocean evaporation, and related surface meteorological variables. Woods Hole Oceanographic Institution OAFlux Project Tech. Rep. Technical Report. OAFlux Project Tech. Rep. OA-2008-01.

Zweng, M., Seidov, D., Boyer, T., Locarnini, M., Garcia, H., Mishonov, A., Baranova, O., Weathers, K., Paver, C., Smolyar, I., et al., 2019. World ocean atlas 2018, volume 2: Salinity .